\documentstyle{pasj06}
\begin{document}

\SetRunningHead{A.Imada et al.}{dwarf nova FL TrA and CTCV J0549-4921}

\title{Photometric Studies of New Southern SU UMa-type dwarf novae,
FL Triangulum Australe and CTCV J0549-4921}

\author{
      Akira Imada$^1$,
      Taichi Kato$^1$, 
      L.A.G. Monard$^2$,
      Rod Stubbings$^3$, \\
      Makoto Uemura$^4$,
      Ryoko Ishioka$^5$, and
      Daisaku Nogami$^6$
}

\affil{$^1$ Department of Astronomy,Faculty of Science, 
       Kyoto University, Sakyo-ku, Kyoto 606-8502, Japan}

\email{a\_imada@kusastro.kyoto-u.ac.jp}

\affil{$^2$ Bronberg Observatory, CBA Pretoria, PO Box 11426,
       Tiegerpoort 0056, South Africa}
       
\affil{$^3$ Tetoora Observatory, Tetoora Road, Victoria, Australia}
       
\affil{$^4$ Hiroshima Astrophysical Science Center, Hiroshima University,
       Hiroshima 739-8526, Japan}

\affil{$^5$ Subaru Telescope, National Astronomical Observatory of Japan
       650 North A'ohoku Place, Hilo, \\HI 96720, U.S.A.}
        
\affil{$^6$ Hida Observatory, Kyoto University, Kamitakara, Gifu
       506-1314, Japan}


\maketitle

\begin{abstract}

We report time-resolved optical CCD photometry on newly discovered SU
 UMa-type dwarf novae, FL TrA and CTCV J0549-4921. During the 2006 August
 outburst, we detected superhumps with a period of 0.59897(11) days for FL
 TrA, clarifying the SU UMa nature of the system. On the first
 night of our observations on FL TrA, the object showed no superhumps.
 This implies that it takes a few days for full development of
 superhumps. The superhump period variation diagram of FL TrA was
 similar to that observed in some WZ Sge stars and short period SU
 UMa-type stars. This indicates that the system is closely related to WZ
 Sge stars and SU UMa stars having short orbital periods. For CTCV
 J0549-4921, the candidates of the mean superhump period are
 0.083249(10) days and 0.084257(8) days, respectively. Due to a lack of
 the observations, we cannot determine the true superhump period, but
 the latter period is favorable. Using the ASAS-3 archive, it
 turned out that the system shows only four outbursts over the past 6
 years. The outburst amplitude of CTCV J0549-4921 was relatively small,
 with about 4.5 mag. One possibility is that mass evaporation may
 play a role during quiescence.
   
\end{abstract}

\section{Introduction}

Dwarf novae are a subclass of cataclysmic variables that consist of a
white dwarf (primary) and a late-type star (secondary). The secondary
star fills its Roche lobe, transferring the matter into the primary via
inner Lagragian point (L1). Then the accretion disc is formed around the
white dwarf. The accretion disc shows various modulations both in
outburst and quiescence (for a review, see \cite{war95book};
\cite{hel01book}; \cite{las01DIDNXT}; \cite{smi07review}).

SU UMa-type stars are a subclass of dwarf novae (\cite{osa89suuma};
\cite{osa96review}; \cite{pat05suuma}; \citet{2005PJAB...81..291O}). The
systems basically exhibit two types of eruptions: normal outburst which
lasts a few days and superoutburst which lasts about two
weeks. During the superoutburst, modulations having an
amplitude of ${\sim}$ 0.2 mag called
superhumps, are always observed. The period of the superhumps are a
few percent longer than that of the orbital period of the system, which
is attributed to prograde precession of tidally deformed accretion disc
\citep{whi88tidal}. Short and long-term variations of SU UMa stars are
well reproduced by the thermal-tidal instability
model developed by \citet{osa89suuma}. Recent arising problems concerning
SU UMa stars are reviewed in \citet{nog07suuma}. 

Recently, the advent of the Internet has significantly improved our
understanding in SU UMa stars (\cite{pat00iyuma}; \cite{pat02wzsge};
\cite{ish02wzsgeletter}; \cite{pat03suumae}; \cite{kat04vsnet}),
especially in the Northern hemisphere. As for Southern SU UMa stars, it
is true that a lot of studies have been performed for them including
e.g., VW Hyi (\cite{vog74vwhyi}; \cite{hae79lateSH}), Z Cha
(\cite{woo86zcha}; \cite{wad88zcha}), and OY Car (\cite{woo89oycar};
\cite{hor94oycarHST}). However, there are many poorly studied SU UMa
stars compared to the Northern SU UMa stars. This trend has been
changing over the past few years because of the advent of the All Sky
Automated Survey (ASAS, \cite{poj02asas3}). Valuable observations of SU
UMa stars have been carried out thanks to prompt detection of outburst by
the ASAS-3 (\cite{tem06asas0025}; \cite{ima06asas1600letter};
\cite{ima062qz0219}).

In this paper, we report photometric observations of two Southern dwarf
novae, FL TrA and CTCV J0549-4921 during outbursts, during which we
detected superhumps for the first time for these objects.

\section{FL TrA}

\subsection{introduction}

FL TrA was first cataloged in \citet{mei70fltra} in which the system was
numbered S 5770 TrA with the variable type of UG. \citet{DownesCVatlas1}
tabulated cataclysmic variable stars,
including FL TrA, in which the variable was categorised as UG with the
magnitude range of 15.5p $-$ 17.0p. \citet{DownesCVatlas1} also gave the
coordinate of RA:$16^{\rm h} 30^{\rm m} 37^{\rm s}$, Dec:$-61^{\circ} 50'
33''$. No outburst of FL TrA was reported to the VSNET
\citep{kat04vsnet} until 2005. We have suspected the WZ Sge subclass of
the object. Spectroscopic observations were carried out by
\citet{mas03faintCV} in which the object showed a spectrum of a common
G-type star. \citet{mas03faintCV} pointed out the misidentification of
FL TrA.

On 2005 July 27, Rod Stubbings reported to the VSNET that FL TrA
appeared to be in outburst with a visual magnitude of 15.0
([vsnet-alert 8574]). He further noticed that the position of
the system looked slightly north from the above mentioned
coordinate. It turned out that FL TrA was misidentified as USNOB1
0281-0691553 (RA:$16^{\rm h} 30^{\rm m} 36^{\rm s}.4$, Dec:$-61^{\circ} 50'
28''.1$). In response to the report, \citet{DownesCVatlas3} refined the
finding chart of the system, which can be seen from the
website.\footnote{$\langle$http://archive.stsci.edu/prepds/cvcat/index.html$\rangle$}
The precise coordinate of the system is RA:$16^{\rm h} 30^{\rm m} 36^{\rm
s}.61$, Dec:$-61^{\circ} 50' 21''.0$, where no optical counterpart
exists in the USNO B1 catalog, which indicates the magnitude in quiescence
may be fainter than 21 mag.

\subsection{observations}

\begin{table}
\caption{Observation log of FL TrA during the 2005 August superoutburst.}
\begin{center}
\begin{tabular}{cccc}
\hline\hline
2005 Date & Start$^a$ & End$^a$ & N$^b$ \\ 
\hline
Jul. 27 & 579.2491 & 579.3572 & 225 \\
Jul. 28 & 580.2378 & 580.5585 & 451 \\
Jul. 29 & 581.2032 & 581.5047 & 426 \\
Jul. 30 & 582.3141 & 582.5241 & 297 \\
Jul. 31 & 583.2259 & 583.4713 & 347 \\
Aug. 1. & 584.1918 & 584.4763 & 360 \\
Aug. 2. & 585.2243 & 585.3945 & 241 \\
\hline
\multicolumn{4}{l}{$^a$ HJD - 2453000. $^b$ Number of exposure.} \\
\end{tabular}
\end{center}
\label{t1}
\end{table}

Time resolved CCD photometric observations were carried out from 2005
July 27 to 2005 August 2 at Bronberg Observatory in South Africa using
a 32 cm Schmidt-Cassegrain telescope equipped with a SBIG ST-7XME CCD
camera. We tabulate journal of observations
in table 1. All of the observations were performed with 30 sec
exposure time. The total data points amounted to 2347. No filter was used
during the observations. The unfiltered data are close to the $R_{\rm
c}$ system. After debiasing and flat-fielding, we performed aperture
photometry using AIP4WIN software. As a comparison star, we used USNO A2.0
0225$-$25536030 (RA:$16^{\rm h} 30^{\rm m} 38^{\rm
s}.79$, Dec:$-61^{\circ} 49' 58''.0$, $B$ =14.2, $R$ = 13.2), whose
constancy was checked by some stars located in the same
image. The 1-sigma error for each differential magnitude is of an order of
0.03 mag, which is small enough to perform the following analysis,
including exploring superhump period and profile
variations. Heliocentric corrections to our run were applied before the
following analysis.

\subsection{results}


\begin{figure}
\begin{center}
\resizebox{80mm}{!}{\includegraphics{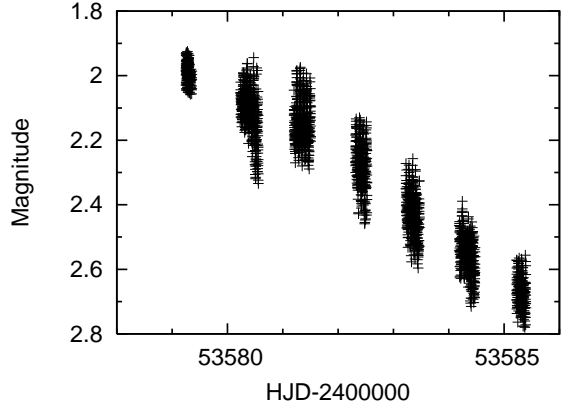}}
\end{center}
\caption{Light curves of FL TrA during the 2005 July/August
 superoutburst. The vertical and horizontal axis indicate differential
 magnitude and the fractional HJD, respectively. The magnitude of the
 comparison star is 13.2 in $R$. The star showed almost the constant
 decline from HJD 2453581 (2005 July 28) at a rate of 0.13 mag
 d$^{-1}$.}
\label{}
\end{figure}

\begin{figure}
\begin{center}
\resizebox{80mm}{!}{\includegraphics{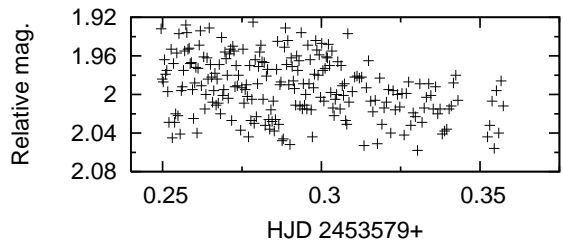}}
\end{center}
\caption{Enlarged light curve on HJD 2453579 (2005 July 27), the first
 night of our run. The light curves provide no evidence of superhumps
 during this phase.}
\label{}
\end{figure}

\begin{figure}
\begin{center}
\resizebox{80mm}{!}{\includegraphics{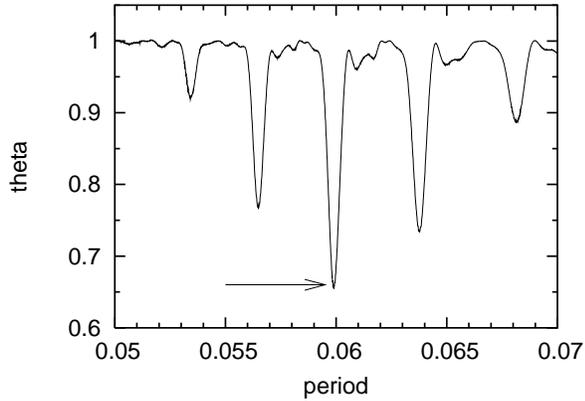}}
\end{center}
\caption{Theta diagram of the superoutburst of FL TrA from HJD 2453580
 to HJD 2453585. The arrow shows the best estimated superhump period,
 0.059897(11) days with 99$\%$ significance level.}
\label{}
\end{figure}

Figure 1 shows light curves of FL TrA during the 2005 July/August
superoutburst. At the onset of our observations, FL TrA was at the
magnitude of 15.0 on 2005 July 27, after which the system almost
constantly declined at the rate of 0.13(1) mag d$^{-1}$. This decline rate
is a typical value among SU UMa-type dwarf novae. As can be seen in
figure 2, the light curve showed almost no feature on 2005 July 27 (HJD
2453579), indicating that superhumps did not yet develop. After subtracting
a linear decline trend of daily light curves, we performed a period
analysis of the phase
dispersion minimization (PDM) method \citep{ste78pdm} applied between HJD
2453580 and 2453285. Figure 3 displays the results of the PDM analysis,
by which we determined 0.059897(11) days as the best estimated period
during this stage. A statistical F-test provided the confidence level of
99 $\%$. The 1-sigma error was calculated using the Lafler-Kinman method
\citep{fer89error}.

\begin{figure}
\begin{center}
\resizebox{80mm}{!}{\includegraphics{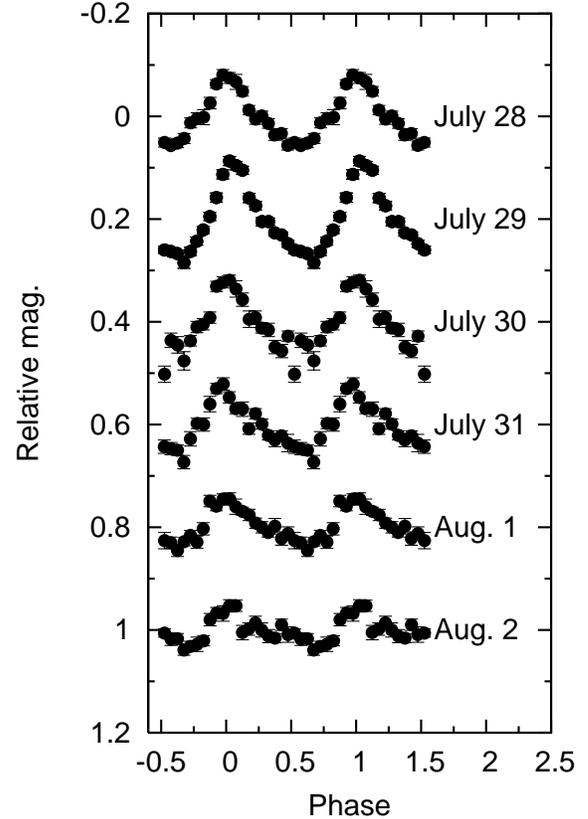}}
\end{center}
\caption{Phase averaged light curve during the superoutburst folded by
 0.059897 days. The abscissa and ordinate denote phase and differential
 magnitude, respectively. A rapid rise and slow decline, characteristic
 of superhumps, are visible.}
\label{}
\end{figure}

We present daily averaged light curves in figure 4. These light curves
are folded
with the above obtained period. On HJD 2453580 (2005 July 28), the
profile is characteristic of superhumps, with the mean amplitude of about
0.2 mag, from which we first confirmed the SU UMa nature of FL TrA. On
2005 July 29, the amplitude of the superhumps was at the maximum value
of about 0.3 mag. No eclipse feature was detected during the
observations, indicating a low-to-mid inclination of FL TrA.

\begin{table}
\caption{superhump timing maxima}
\begin{center}
\begin{tabular}{ccc}
\hline\hline
$E^a$ & Time$^b$ & error$^c$ \\
\hline
0 & 580.2921 & 0.003 \\
1 & 580.3506 & 0.004 \\
2 & 580.4116 & 0.002 \\
3 & 580.4712 & 0.002 \\
4 & 580.5355 & 0.003 \\
16 & 581.2542 & 0.002 \\
17 & 581.3122 & 0.004 \\
18 & 581.3702 & 0.001 \\
19 & 581.4297 & 0.001 \\
20 & 581.4920 & 0.003 \\
34 & 582.3274 & 0.001 \\
35 & 582.3879 & 0.003 \\
36 & 582.4532 & 0.004 \\
50 & 583.2827 & 0.002 \\
51 & 583.3451 & 0.004 \\
52 & 583.4002 & 0.002 \\
66 & 584.2420 & 0.002 \\
68 & 584.3643 & 0.005 \\
69 & 584.4266 & 0.003 \\
83 & 585.2644 & 0.001 \\
84 & 585.3212 & 0.004 \\
\hline
\multicolumn{3}{l}{$^a$ Cycle count. $^b$ HJD - 2453000.} \\
\multicolumn{3}{l}{$^c$ error in a unit of days.} \\
\end{tabular}
\end{center}
\end{table}

\begin{figure}
\begin{center}
\resizebox{80mm}{!}{\includegraphics{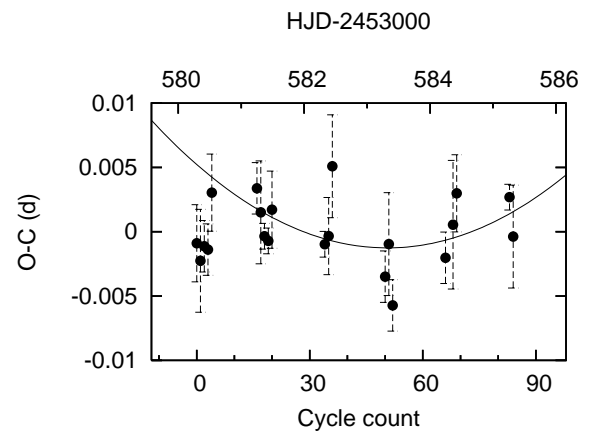}}
\end{center}
\caption{$O-C$ diagram of FL TrA. Each datapoint of the maximum timing
 of superhumps is listed in table 2. The solid curve means the best
 fitting quadratic described in equation (2) for 16 $<E<$ 84. It
 should be noted that the cycle count between 0 $<$ E $<$ 3 is
 deviated from the quadratic, which is also observed in some short
 period SU UMa stars.}
\end{figure}

In order to investigate the variations of the superhump period during
the plateau phase, we measured the timings of the superhump maxima
listed in table 2. The typical error is an order of 0.002 days. A linear
fitting yielded as the following equation, 

\begin{equation} 
HJD (max) = 2453580.2930(9) + 0.059864(21) \times E,
\end{equation}

where the parentheses denote 1-sigma error for each value. By using the
above ephemeris, we draw an $O - C$ diagram, which is displayed in
figure 5. The best fitted quadratic for 16 $< E <$ 84 can be
represented as follows:

\begin{eqnarray}
O - C =&  5.23(2.98)\times10^{-3} - 2.56(1.46)\times10^{-4} E \nonumber \\
       & +2.53(1.50)\times 10^{-6} E^{2}.
\end{eqnarray}

The above obtained value implies that the superhump period may increase since
HJD 2453581 with $P_{\rm dot}$ = $\dot{P}$/$P$ =
$+$8.4(5.0)$\times$10$^{-5}$.

\subsection{FL TrA as a short period SU UMa star}

The present photometric studies and the previous archival survey
reasonably qualified FL TrA as a new member of SU UMa-type dwarf novae
with short periods. The outburst amplitude of FL TrA exceeded 6
mag, suggestive of a large amplitude SU UMa-type dwarf novae (TOAD,
\cite{how95TOAD}). Unfortunately, the lack of baseline during the early
stage of the superoutburst prevented us from further investigating whether
double-peaked humps existed, which is exclusively observed among WZ Sge
stars in early phase of superoutburst (\cite{osa02wzsgehump};
\cite{kat02wzsgeESH}; \cite{pat02wzsge}). 

As for superhump period changes, the estimated positive $P_{\rm dot}$
derivative indicates that the
superhump period increases during the plateau stage. Such systems
include all of
confirmed WZ Sge-type dwarf novae \citep{kat01hvvir}, as well as SU
UMa-type dwarf novae with short superhump periods
\citep{oiz07v844her}. Recently,
\citet{uem05tvcrv} found that a short period SU UMa star TV Crv
shows two types of $P_{\rm dot}$. The 2001 superoutburst of TV Crv
showed positive $P_{\rm dot}$, while the 2004 superoutburst showed almost
constant $P_{\rm sh}$. The big difference between the two
superoutbursts is not only the different $P_{\rm dot}$, but also the
light curves themselves: a precursor was present for the 2004
superoutburst while it was absent for the 2001 superoutburst. One
interpretation is that an appearance of the positive or
constant/negative $P_{\rm dot}$ depends
on the maximum radius of the accretion disc during the superoutburst
\citep{uem05tvcrv}. \citet{uem05tvcrv} have further stated that systems
which show both types of $P_{\rm dot}$ will be restricted to
short period SU UMa stars, because the tidal truncation radius should be
significantly larger than the 3:1 resonance radius (see
\cite{osa03DNoutburst}). With this respect, the 2005 superoutburst of FL
TrA had a large disc radius, which is consistent with the large
amplitude of the outburst.

Another important finding is that unfittable cycle counts by the
equation (2) exists at the earliest stage of our run. These correspond to
0 $< E <$ 4 in figure 5. Similar results are found in V1028 Cyg
\citep{bab00v1028cyg}, RZ Leo \citep{ish01rzleo}, HV Vir
\citep{ish03hvvir}, V844 Her \citep{oiz07v844her} and GW Lib (Imada et
al. in preparation). During the cycle count of 0 $< E <$ 4, the
superhump period keeps constant, while an abrupt change of the superhump
period occurred after $E >$ 4. The origin of the abrupt period
change remained unknown, which should be elucidated in the future
observations.

\section{CTCV J0549-4921}

\subsection{introduction}

CTCV J0549-4921 (hereafter CTCV J0549) was first identified as a
candidate of cataclysmic
variables after spectroscopic observations in the Calan-Tololo Survey
\citep{maz89CTS}. The optical spectrum shows H${\alpha}$ and HeI 5876
emission \citep{tap04ctcv}, indicating the dwarf nova nature of the system.
\citet{tap04ctcv} pointed out there is no evidence of the secondary
star in the optical spectrum because of the absence of TiO bands. Optical
observations during quiescence and outburst were also performed by
\citet{tap04ctcv}. During the quiescence, \citet{tap04ctcv} found
photometric orbital
modulations with the period of 0.080218(70) days, which they interpreted
as the orbital period of CTCV J0549. \citet{tap04ctcv} pointed out the
shape of the modulation is reminiscent of quiescent light curve of WZ
Sge. During the outburst,
CTCV J0549 brightened up to $V$ = 13.75 on 1996 October 5. However,
superhumps were not detected. It is likely
\citet{tap04ctcv} observed a normal outburst of SU UMa-type dwarf
novae. In conjunction with the above observations, CTCV J0549 has been a
promising candidate for SU UMa-type dwarf novae.

On 2006 April 2, a brightening of the star was discovered by
L.A.G. Monard ([vsnet-alert 8896]) at the magnitude of 13.8, who
detected the rising phase of outburst. On 2006 April 4, we first
detected superhumps of CTCV J0549, and confirmed the SU UMa nature of
the object. Long-term monitoring by
the ASAS-3 have detected 3 outbursts, of which one was possibly a
superoutburst. This occurred in 2004 January. The precise coordinate of
the system is RA:$05^{\rm h} 49^{\rm m} 45^{\rm s}.4$, Dec:$-49^{\circ}
21' 56''$, where the 2MASS counterpart of CTCV J0549 yields $J$ =
15.619(50), $H$ = 15.210(81), and $K$ = 14.869(112), respectively
\citep{ima06j0137}.

\subsection{observations}

\begin{table}
\caption{Observation log of CTCV J0549 during the 2006 April superoutburst.}
\begin{center}
\begin{tabular}{cccc}
\hline\hline
2006 Date & Start$^a$ & End$^a$ & N$^b$ \\
\hline
Apr. 2 & 828.2200 & 828.3839 & 452 \\
Apr. 4 & 830.2275 & 830.3658 & 286 \\
Apr. 5 & 831.1955 & 831.3463 & 428 \\
Apr. 6 & 832.1991 & 832.3524 & 428 \\
Apr. 12& 838.1970 & 838.3339 & 385 \\
\hline
\multicolumn{4}{l}{$^a$ HJD - 2453000. $^b$ Number of exposure.}
\end{tabular}
\end{center}
\end{table}

Time resolved CCD photometric observations were carried out from 2006
April 2 to 2006 April 12. The observing site and instrument are the same
as described in section 2.2. The journal of observations is summarized in
table 3. All of the observations were performed with 30-sec exposure time
with no filter. The total data points of our run amounted to 1979. For
obtained data, we performed the same manner as mentioned in section
2.2. We used  USNO A2.0 0375$-$2158238 
(RA:$05^{\rm h} 49^{\rm m} 53^{\rm s}.41$,
Dec:$-49^{\circ} 18' 50''.8$, $B$ = 13.3, $R$ = 12.8) as a
comparison star, whose constancy was checked by some stars in the same
image. The 1-sigma error for each differential magnitude is of an order of
0.01 mag. Heliocentric corrections to our run were applied before the
following analyses. 

\subsection{results}

\begin{figure}
\begin{center}
\resizebox{80mm}{!}{\includegraphics{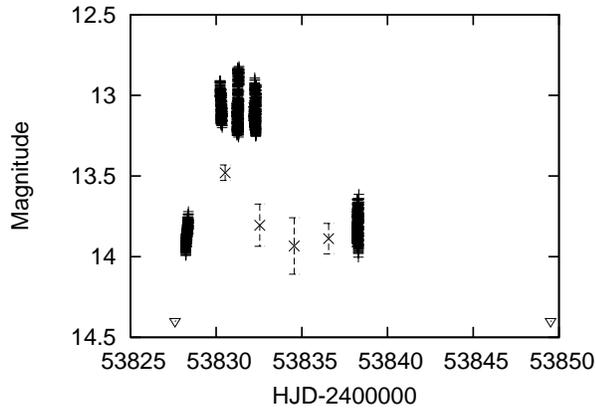}}
\label{}
\end{center}
\caption{Light curves of CTCV J0549 during the 2006 April
 superoutburst. The ordinate means the ASAS-3 $V$ and $R$ magnitude. The
 magnitude of the comparison star is 12.7 in $R$. The filled circles
 show time resolved CCD observations. The crosses indicate the ASAS-3
 light curves, which contains 0.2 mag error originated from the
 modulation of superhumps. The negative observation was performed by
 the ASAS-3 on HJD 2453849, when the object was fainter than 14.4 in
 $V$.}
\end{figure}

The overall light curves during our run are presented in figure 6, in
which we also demonstrate the ASAS-3 positive and negative
observations. The discrepancy between our CCD observations and
the ASAS-3 archive is large, presumably due to different filters between
the site and the ``snapshot'' in the ASAS-3 photometry. Nevertheless,
it is well determined that the bright maximum of CTCV J0549 was on HJD
2453831 with the magnitude of ${\sim}$ 13.0. The rarely observed rising
phase was fortunately detected on HJD 2453879, providing $-$1.0(1) mag
d$^{-1}$ as the rising rate. Although our observations were absent
between HJD 2453833-2453837, during which the magnitude was assumed to
decline linearly, the decline rate could be estimated to be 0.12(1)
mag $^{-1}$. The value is typical for usual SU UMa-type dwarf novae
during the plateau phase. 

\begin{figure}
\begin{center}
\resizebox{80mm}{!}{\includegraphics{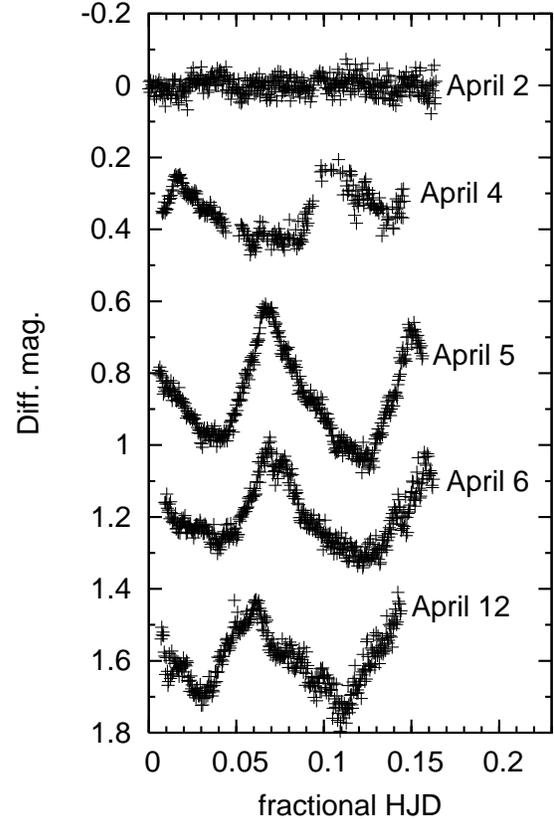}}
\label{}
\end{center}
\caption{Daily light curves after removing the linear trend. There were
 almost no modulation on April 2. The growth time of superhumps is as
 short as 2 days. The amplitude of superhumps is the largest on April 5
 with 0.4 mag. There is no evidence for an eclipse, which indicates of
 low to mid inclination system.}
\end{figure}

Enlarged light curves for each night are depicted in figure 7, after
subtracting linear rising or declining trend. As can be seen in this
figure, there are no features on HJD 2453829 (2006 April 2), corresponding to
the rising phase, while prominent superhumps are shown from HJD 2453831
(2006 April 4). Therefore, we first confirmed CTCV J0549 as an SU UMa
star.

\begin{table}
\caption{superhump timing maxima}
\begin{center}
\begin{tabular}{ccc}
\hline\hline
$E^a$ & Time$^b$ & error$^c$ \\
\hline
0 & 830.2360 & 0.001 \\
1 & 830.3230 & 0.005 \\
12 & 831.2567 & 0.002 \\
13 & 831.3401 & 0.001 \\
24 & 832.2591 & 0.001 \\
25 & 832.3474 & 0.001 \\
\hline
\multicolumn{3}{l}{$^a$ Cycle count. $^b$ HJD - 2453000.} \\
\multicolumn{3}{l}{$^c$ error in unit of days.} \\
\end{tabular}
\end{center}
\end{table}

\begin{figure}
\begin{center}
\resizebox{80mm}{!}{\includegraphics{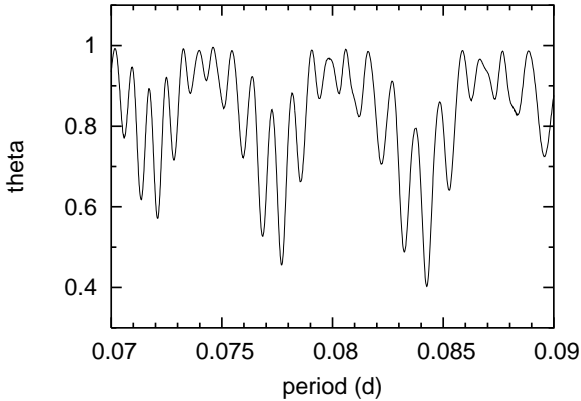}}
\end{center}
\caption{Theta diagram of CTCV J0549 applied to the plateau phase. Two
 possible periods of superhumps were found, 0.083249(10) days and
 0.084257(8) days. Due to the lack of the observations, we
 cannot specify the exact period of superhumps.}
\end{figure}

In order to determine the mean superhump period during the plateau
phase, we performed the PDM method \citep{ste78pdm}. The strongest
periodicity can be found at 0.084257(8) days. However, due to the lack of
our observations, we cannot rule out the second strongest period,
0.083249(10) days. Hence, we carried out another approach to
determine the mean superhump period by measuring the maximum timing of
the superhumps. We tabulate the result on table 4. The best fitting
linear regression is yielded in the following equation:

\begin{equation}
HJD(max) = 2453830.2396(29) + 0.08433(18) \times E.
\end{equation}

The above equation favors the former period of the superhump, 0.084257(8)
days. If the quiescent modulations reflect the orbital period of the
system, and the mean superhump period is $P_{\rm sh}$=0.084257(8) days,
then the fractional superhump period excess is ${\sim}$ 5 $\%$. This
value is significantly larger than that observed in common SU UMa-type
dwarf novae \citep{pat05suuma}. The actual $P_{\rm orb}$ and $P_{\rm
sh}$ should be measured in the future observations. 

\subsection{CTCV J0549 as a long period SU UMa star}

\begin{table}
\caption{Recorded outbursts by the ASAS-3.}
\begin{center}
\begin{tabular}{cccc}
\hline\hline
Time$^a$ & Mag.$^b$ & error$^c$ & type$^d$\\
\hline
51949.60890 & $<$14.4 & - & \\
51952.54990 & 13.515 & 0.062 & N \\
51954.55382 & $<$14.4 & - & \\
52171.80528 & $<$14.4 & - & \\
52172.86305 & 13.654 & 0.047 & ? \\ 
52183.79372 & $<$14.4 & - & \\
53020.83195 & $<$14.4 & - & \\
53025.61437 & 13.714 & 0.023 & S \\ 
53029.61569 & 13.284 & 0.067 & \\
53031.65788 & 13.732 & 0.111 & \\
53033.68195 & 14.580 & 0.437 & \\
53035.64670 & 14.006 & 0.075 & \\
53039.64744 & $<$14.4 & - & \\
53827.59870 & $<$14.4 & - & \\
53830.53230 & 13.480 & 0.048 & S \\ 
53832.54772 & 13.805 & 0.131 & \\
53834.55680 & 13.934 & 0.175 & \\
53836.56170 & 13.888 & 0.094 & \\
53849.51095 & $<$14.4 & - & \\
\hline
\multicolumn{4}{l}{$^a$ HJD - 2400000. $^b$ Mean magnitude in} \\
\multicolumn{4}{l}{$V$. $^c$ 1-sigma error in unit of $V$.} \\
\multicolumn{4}{l}{$^d$N: Normal outburst. S: Superoutburst.} \\
\end{tabular}
\end{center}
\end{table}

We first confirmed the SU UMa nature of CTCV J0549 by the detection of
superhumps. Although the mean superhump period cannot be determined, the
period exceeds 0.08 days, which we safely qualify CTCV J0549 as a long
period SU UMa star. This is also supported by quiescent photometric
observations \citep{tap04ctcv}. 

The most remarkable fact for CTCV J0549 is that the object has shown only
4 outbursts over the past 6 years. According to the ASAS-3 archive, the
recorded outbursts were 2001 Febuary 12, 2001 September 20, 2004 January
21, and the present superoutburst. We summarize recorded outbursts
monitored by the ASAS-3 in table 5. Judging from table 5, only two
are superoutbursts, one is a normal outburst, and we cannot distinguish
the type for one outburst. If we do not miss any superoutburst since
2001, a supercycle of CTCV J0549 is estimated as ${\sim}$ 800
days. This is one of the longset values among SU UMa-type dwarf novae
\citep{kat01hvvir}. Inactive systems
having a similar superhump period include QY Per ($P_{\rm sh}$ = 0.07681
days, \cite{kat00qyperiauc}), EF Peg ($P_{\rm sh}$ = 0.08705 days,
\cite{kat02efpeg}) and V725 Aql ($P_{\rm sh}$ = 0.09909 days,
\citep{uem01v725aql}). Although the exact mechanism of the long
supercycle still remains unknown, mass evaporation
during quiescence might be a possible explanation for the origin of the
outburst and quiescent properties (\cite{mey94siphonflow}; \cite{las95wzsge};
\cite{min98wzsge}). This may be consistent with relatively small
amplitude of 4.5 mag of CTCV J0549. As for evporation,
\citet{min98wzsge} predicted
that quiescent superhumps could be observed even during quiescence if
evaporation works in the acretion disc. \citet{oiz07v844her} also argued
that a peak separation variation of an optical spectrum during
quiescence is a powerful tool to check whether or not the evaporation
works in the accretion disc. Future spectroscopic observations are
required to elucidate the nature of CTCV J0549. 

\section{Summary}

In this paper, we newly confirmed the SU UMa nature of FL TrA and CTCV
0549-4921.  

After the discovery of the outburst of FL TrA, we found that the
previous candidate of the object had been misidentified. The mean
superhump period of FL TrA was determined to be 0.059897 days. This
superhump period qualified FL TrA as a short period SU UMa-type dwarf
nova. The superhump period
increased at the rate of $P_{\rm dot}$ = $+$8.4(5.0)$\times$10$^{-5}$,
which is a typical value of short period SU UMa stars. At the early
stage of the superoutburst, the period of the superhumps changed
abruptly, as observed in some WZ Sge stars, as well as short period SU
UMa stars. Although the exact mechanism of the abrupt change is unknown,
the origin of this phenomenon should be discussed in future.

A previously suspected dwarf nova CTCV J0549-4921 has been first
confirmed as the SU UMa nature by the detection of
superhumps. Although a short baseline of our observation hindered us from
accurate determination of the mean superhump period, we found the
strongest signal at 0.084257 days, which is consistent with our eye
estimation. This candidate period leads us to the conclusion that CTCV
J0549 belongs to a long period SU UMa star. The ASAS-3 archive for CTCV
J0549 puzzles us in terms of its inactive behaviour, as well as the small
amplitude of the outburst despite the long supercycle of the system. A
possible explanation may be that the mass evaporation plays a role
during quiescence.

\vskip 5mm

We would express our gratitude to G. Pojmanski for
providing invaluable data of ASAS-3 observations. This work is supported
by a Grant-in-Aid for the 21st Century COE ``Center for Diversity and
Universality in Physics'' from the Ministry of Education, Culture,
Sports, Science and Technology (MEXT). This work is partly supported by
a grant-in aid from the Ministry of Education, Culture, Sports,
Science and Technology (No. 17740105). Part of this work is
supported by a Research Fellowship of the Japan Society for the
Promotion of Science for Young Scientists (RI, AI).

\end{document}